\documentclass[aps,prb,reprint,showpacs,superscriptaddress,longbibliography,floatfix]{revtex4-2}
\usepackage{mathtools,amssymb,graphicx,units}
\usepackage{amsmath,bm}
\usepackage{bbold}

\usepackage{multirow}

\usepackage[plainpages=false,pdfpagelabels,colorlinks=true,linkcolor=red,urlcolor=blue,citecolor=blue,pdftitle={Title},pdfauthor={},pdfdisplaydoctitle=true,pdfduplex=DuplexFlipLongEdge]{hyperref}
\usepackage{verbatim}
\usepackage[english]{babel}
\usepackage{xcolor}
\usepackage{dsfont}

\usepackage{threeparttable}
\usepackage{tablefootnote}
\usepackage{fixfoot}

\usepackage[normalem]{ulem}

\usepackage{braket}

\usepackage{tikz}
\usetikzlibrary{topaths}

\newcommand{\tcr}{\textcolor{red!85!black}}
\newcommand{\tcb}{\textcolor{blue}}
\newcommand{\tcg}{\textcolor{green!45!black}}

\newcommand\sblacksquare[1][.5]{\mathbin{\vcenter{\hbox{\scalebox{#1}{$\blacksquare$}}}}}

\newcommand\sBox[1][.5]{\mathbin{\vcenter{\hbox{\scalebox{#1}{$\Box$}}}}}

\newcommand\sbullet[1][.5]{\mathbin{\vcenter{\hbox{\scalebox{#1}{$\bullet$}}}}}

\newcommand\scirc[1][.5]{\mathbin{\vcenter{\hbox{\scalebox{#1}{$\circ$}}}}}

\newcommand\sblacktriangleright[1][.5]{\mathbin{\vcenter{\hbox{\scalebox{#1}{$\blacktriangleright$}}}}}
\newcommand\svartriangleright[1][.5]{\mathbin{\vcenter{\hbox{\scalebox{#1}{$\vartriangleright$}}}}}

\newcommand\sblacklozenge[1][.5]{\mathbin{\vcenter{\hbox{\scalebox{#1}{$\blacklozenge$}}}}}

\newcommand{\ketbra}[2]{\mathinner{|{#1}\rangle \langle{#2}|}}
\allowdisplaybreaks

\begin{document}

\title{Altermagnetism and beyond in the $t$-$t^\prime$-$\delta$ Fermi-Hubbard model}

\author{Saisai He}
\affiliation{Lanzhou Center for Theoretical Physics, Lanzhou University, Lanzhou 730000, China}
\affiliation{Key Laboratory of Quantum Theory and Applications of MoE, Lanzhou University, Lanzhou 730000, China}
\affiliation{Key Laboratory of Theoretical Physics of Gansu Province$\&$Gansu Provincial Research Center for Basic Disciplines of Quantum Physics, Lanzhou University, Lanzhou 730000, China}

\author{Jize Zhao}
\email[]{zhaojz@lzu.edu.cn}
\affiliation{Lanzhou Center for Theoretical Physics, Lanzhou University, Lanzhou 730000, China}
\affiliation{Key Laboratory of Quantum Theory and Applications of MoE, Lanzhou University, Lanzhou 730000, China}
\affiliation{Key Laboratory of Theoretical Physics of Gansu Province$\&$Gansu Provincial Research Center for Basic Disciplines of Quantum Physics, Lanzhou University, Lanzhou 730000, China}

\author{Hong-Gang Luo}
\affiliation{Lanzhou Center for Theoretical Physics, Lanzhou University, Lanzhou 730000, China}
\affiliation{Key Laboratory of Quantum Theory and Applications of MoE, Lanzhou University, Lanzhou 730000, China}
\affiliation{Key Laboratory of Theoretical Physics of Gansu Province$\&$Gansu Provincial Research Center for Basic Disciplines of Quantum Physics, Lanzhou University, Lanzhou 730000, China}

\author{Shijie Hu}
\email[]{shijiehu@csrc.ac.cn}
\affiliation{Beijing Computational Science Research Center, Beijing 100084, China}
\affiliation{Department of Physics, Beijing Normal University, Beijing, 100875, China}

\begin{abstract}
In this work, we revisit the phase diagram of the $t$-$t^\prime$-$\delta$ Fermi-Hubbard model on the square lattice to gain a more comprehensive understanding of this correlated model at half filling.
This model has recently become a prominent topic of research because it hosts altermagnetic phases.
Using mean-field analysis, we identify four metallic phases and two insulating phases with nontrivial magnetic orders at an intermediate value of $\delta = 0.5$, presenting a rich ground-state phase diagram in the $U$-$t^\prime$ plane.
We also highlight the distinct features of the Fermi surface topology for each metallic phase.
To go beyond the mean-field theory, we employ the density-matrix renormalization group method to simulate the ground state numerically.
The phase boundaries are determined from the discontinuities and peaks in the entanglement entropy and magnetizations. In addition to the phases identified in the mean-field theory, we find a valence-bond solid state in a narrow intermediate-$t'$ region.
Our work offers a firm step forward in understanding the complex behaviors of correlated electrons in the $t$-$t^\prime$-$\delta$ Hubbard model over a large parameter space.
\end{abstract}
\maketitle

\section{Introduction}\label{sec:intro}
Magnetism is ubiquitous in nature and it may be one of the most extensively studied phenomena, both theoretically and technically, with wide applications in condensed-matter physics.
Ferromagnets and antiferromagnets are two typical representatives in this area, with the spins aligned parallel in the former and antiparallel in the latter.
Due to the protection by translation and/or inversion symmetry, the energy bands in antiferromagnets are spin-independent, making them insensitive to external probes.
While antiferromagnets may host many exotic physical phenomena, it is difficult to find their applications in spintronics in comparison with the ferromagnets.
Recently, altermagnets~\cite{PhysRevX.12.040002, PhysRevX.12.031042, PhysRevX.12.040501}, recognized as the third category of collinear magnets, have ignited massive interest 
in the search for magnetic materials with spin-splitting energy bands~\cite{doi:10.7566/JPSJ.88.123702, PhysRevB.102.014422}. 
Although the spin configuration of altermagnets in real space is analogous to that of antiferromagnets, 
their Kramer degeneracy of energy bands is lifted due to the presence of alternating surroundings, resulting in many spin-polarized effects~\cite{https://doi.org/10.1002/adfm.202409327}, such as the anomalous Hall effect~\cite{doi:10.1126/sciadv.aaz8809, Smejkal2022, PhysRevB.110.094425} and spin currents~\cite{Ma2021}.
These effects make the magnetic moment in altermagnets to be detectable by external probes even in the absence of macroscopic net magnetization, thus enabling the manipulation of the spin degrees of freedom of electrons in spintronic devices.

In the past several years, much effort has been made~\cite{Mazin2023, YU2024100021, Sun2023, Chakraborty2024, gao2024, Liu206702, Yanzhao2024} to search for altermagnetic compounds, particularly those with prominent features such as 
a high N\'{e}el temperature~\cite{Takagi2025} and giant spin splitting~\cite{Ding2024}. On the theoretical hand,
understanding the origin of the altermagnetic phases in these materials necessarily involves microscopic correlated models~\cite{PhysRevLett.132.263402, PhysRevLett.133.086503, PhysRevB.110.205120, PhysRevB.110.205140, giuli2024, delre2024dirac, menghan2024, zhaomm2024, liuYang2024}, in addition to symmetry analysis~\cite{PhysRevX.12.040002}. 
In the correlated metalic phase, the simplest model to be considered may be the $t$-$t^\prime$-$\delta$ Fermi-Hubbard model~\cite{PhysRevLett.132.263402}.
It has already demonstrated~\cite{PhysRevLett.132.263402} by the mean-field (MF) theory that, at half filling, such a model hosts a normal metallic (NM) phase, an altermagnetic metallic (AMM) phase and an altermagnetic insulating (AMI) phase within the parameter window of low $t^\prime$ and low $\delta$.
In particular, when $\delta$ vanishes the AMI phase becomes an antiferromagnetic (AFM) insulating phase.
For a correlated model, in general, we can expect that these conclusions are incomplete, as some properties of such a model, even at $\delta=0$, remain elusive, particularly when doping away from half filling. Even at half filling, as we will show later, the phase diagram is more complicated than previously presented~\cite{PhysRevLett.132.263402}.

\begin{figure*}[t!]
\centering
\includegraphics[width=0.95\linewidth]{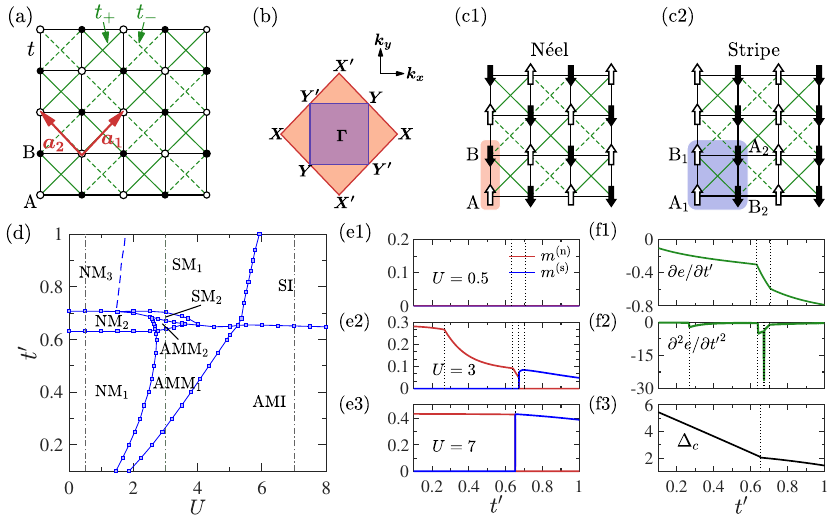}
\caption{(a) The schematic picture marking the model Hamiltonian~\eqref{eq:model}, where black lines represent NN links with hopping coefficient $t$, and green lines represent NNN links with hopping coefficients $t_+$ (solid line) and $t_-$ (dashed line).
On the square lattice with sublattice sites A ($\circ$) and B ($\bullet$), the primitive lattice vectors $\bm{a}_1$ and $\bm{a}_2$ are annotated (see text).
(b) The FBZ (red) for the lattice and the MBZ (blue) for the stripe AFM phase.
In the MF calculations, we choose two distinct ansatzs: local spins are arranged in (c1) the N\'{e}el phase and (c2) the stripe AFM phase, respectively.
In (c1), a pair of A and B sites form a unit cell (red shadow), while in (c2), A\textsubscript{1}, B\textsubscript{1}, A\textsubscript{2} and B\textsubscript{2} are enclosed in a bigger unit cell (blue shadow).
(d) The MF ground-state phase diagram in the TDL features three NM phases (NM\textsubscript{1}, NM\textsubscript{2} and NM\textsubscript{3}), two AMM phases (AMM\textsubscript{1} and AMM\textsubscript{2}), two stripe metallic (SM) phases (SM\textsubscript{1} and SM\textsubscript{2}), an AMI phase, and a stripe insulating (SI) phase.
Along the dash-dotted cutting lines corresponding to on-site repulsion strengths $U=0.5$, $3$ and $U = 7$, we present (e1-e3) the magnetizations $m^{(\text{n})}$, $m^{(\text{s})}$, (f1) the first-order and (f2) the second-order derivatives of the ground-state energy $e$, and (f3) the charge gap $\Delta_\text{c}$ as functions of the hopping coefficient $t^\prime$, respectively.
In (e1-e3,f1-f3), the dotted lines indicate the positions of transition points.
}
\label{fig:fig1}
\end{figure*}

The paper is organized as follows.
In Sec.~\ref{sec:Model}, we introduce the $t$-$t^\prime$-$\delta$ Fermi-Hubbard model on the square lattice and discuss its known features in the two limiting cases of $\delta=0$ and $1$.
In Sec.~\ref{sec:MeanfieldAnalysis}, we present the MF approximation for two specific magnetic structures at an intermediate value of $\delta=0.5$, along with results in the thermodynamic limit (TDL), covering all phases, phase transitions, and the Fermi surface topology.
Section~\ref{sec:DMRGCalculations} presents our main density-matrix renormalization group (DMRG) results for selected torus geometries, focusing on the phase diagram at half filling and comparing it to the MF predictions.
In particular, we identify the emergence of the valence-bond solid~(VBS) state in an intermediate-$t^\prime$ region.
Finally, we summarize our findings in Sec.~\ref{sec:Conclusions}.
In Apps.~\ref{app:appA} and \ref{app:appB}, we provide additional MF results in the TDL and entanglement entropy calculations for other torus geometries for comparison.

\section{Model}\label{sec:Model}

On an $L_x \times L_y$ square lattice composed of sublattice sites A and B, the Hamiltonian of the $t$-$t^\prime$-$\delta$ Fermi-Hubbard model reads
\begin{eqnarray}
\begin{split}
\hat{H} = &- t \sum_{ \braket{l,l'},\sigma} \hat{c}_{l\sigma}^{\dag} \hat{c}_{l'\sigma}^{\phantom{\dag}} - t^\prime \sum_{ \langle\!\langle l,l' \rangle\!\rangle,\sigma} (1 + \text{Sgn}_{ll^\prime} \delta) \hat{c}_{l\sigma}^{\dag} \hat{c}_{l'\sigma}^{\phantom{\dag}}\\
&+ U \sum_l {\hat n}^{\phantom{\dag}}_{l\uparrow} {\hat n}^{\phantom{\dag}}_{l\downarrow}\, ,
\end{split}
\label{eq:model}
\end{eqnarray}
where $\hat{c}_{l\sigma}^{\dag}$ ($\hat c_{l\sigma}^{\phantom{\dag}}$) is the creation (annihilation) operator for an electron with spin $\sigma=\uparrow$, $\downarrow$ at site $l$, $\hat{n}^{\phantom{\dag}}_{l\sigma}=\hat{c}^{\dag}_{l\sigma} {\hat c}^{\phantom{\dag}}_{l\sigma}$ the particle number operator, $t$ and $t^\prime$ the hopping amplitudes, and $U > 0$ the strength of the onsite repulsion.
The index $l$ sums over $2 N = L_x L_y$ sites in $N$ unit cells.
$\braket{l,l'}$ and $\langle\!\langle l,l' \rangle\!\rangle$ runs over all nearest-neighbor (NN) and next-nearest-neighbor(NNN) links, respectively.
The dimensionless parameter $\delta > 0$ represents the relative discrepancy of the hopping amplitudes on the NNN links that reside on two inequivalent plaquettes, corresponding to $\text{Sgn}_{ll'} = \pm 1$. The NNN hoppings can then be written as $t_\pm = t^\prime (1 \pm \delta)$.
For clarity, we refer to the plaquettes with the $t_+$ NNN hoppings as the $t_+$-plaquettes, and the remaining as the $t_-$-plaquettes, as illustrated in Fig.~\ref{fig:fig1}(a).
In the following, $t=1$ is taken as the energy unit.
Our calculations focus on the half filling, meaning that there is, on average, one electron per site.
For zero net magnetization, this condition is represented as $N_{\uparrow} = N_{\downarrow} = N$, where $N_{\sigma} = \sum_{l} \braket{\hat{n}^{\phantom{\dag}}_{l\sigma}}$ denotes the number of electrons with spin $\sigma$.

This model~\eqref{eq:model} is mainly characterized by the U($1$)$\times$SU($2$) group for the charge and spin degrees of freedom, separately~\cite{He_2024}.
Moreover, it does not have particle-hole symmetry at half filling due to the presence of a finite $t^\prime$ on the NNN links, as illustrated in Fig.~\ref{fig:fig1}(a).
Lastly, the Hamiltonian is not invariant under a translation by one lattice spacing along either the $x$ or $y$-axis due to the presence of a nonzero $\delta$.
Specifically, it is invariant under the translation by the vectors $\bm{a}_1 = (1,\, 1)$ or $\bm{a}_2 = (-1,\, 1)$ along the plaquette diagonals, as shown in Fig.~\ref{fig:fig1}(a).
Consequently, the first Brillouin zone (FBZ), represented by the red diamond region in Fig.~\ref{fig:fig1}(b), is half the size of the FBZ for the regular square lattice.
When considering the magnetic structure, the magnetic Brillouin zone (MBZ) may differ.
For the N\'{e}el phase depicted in Fig.~\ref{fig:fig1}(c1), which breaks spin rotation symmetry, the assumed staggered collinear spin configuration aligns with the lattice structure, with spin-up on sublattice sites A and spin-down on sublattices sites B.
In contrast, the stripe AFM phase, depicted in Fig.~\ref{fig:fig1}(c2), features a bigger unit cell that includes a pair of spin-up at sites A\textsubscript{1} and B\textsubscript{1}, and a pair of spin-down at sites A\textsubscript{2} and B\textsubscript{2}.
As a result, the MBZ is only half the size of the FBZ, as indicated by the small blue square region in Fig.~\ref{fig:fig1}(b).

Let us start by outlining some results of the Hamiltonian~\eqref{eq:model} in two specific limits: the uniform limit of $\delta=0$ and the checkerboard limit of $\delta = 1$.

$\bullet$ In the uniform limit, the model~\eqref{eq:model} is just the $t$-$t^\prime$ model.
At $U=0$, it exhibits a Lifshitz transition at $t^\prime / t = 0.5$ associated with a change in the topology of the Fermi surface~\cite{Lifshitz_1960}.
As a finite $U$ is introduced, a collinear stripe AFM order and a possible magnetic order with a longer spatial periodicity arise, competing with the N\'{e}el order in the insulating region~\cite{Yu_2010, Yamada_2013, Timirgazin_2016}.
Recently, advanced numerical simulations have also shown the possible emergence of double-striped-type and triple antiferromagnetic states in the intermediate-$t'$ region~\cite{Zheng_2016, Ruan_2023}.
When $U\gg t$, $t^\prime$, charge degrees of freedom are frozen.
In this regime, second-order hopping processes lead to an effective Heisenberg model, which includes AFM Heisenberg couplings on NN and NNN links, represented as $J_1 = 4t^2/U$ and $J_2 = 4{t^\prime}^2/U$, respectively.
By adjusting the ratio of $J_2 / J_1$, the model reveals a variety of intriguing transition processes.
When $J_2 \ll J_1$, the ground state corresponds to a N\'{e}el phase.
In contrast, when $J_2 \gg J_1$, the ground state favors a four-fold degenerate stripe AFM phase~\cite{Tchernyshyov_2003}, which breaks the lattice $\pi/2$-rotation symmetry~\cite{Hetzel_1998}.
However, the ground states in the intermediate region remain highly controversial, and the promising candidates include a VBS phase, and/or an antiferromagnetic chiral spin density wave phase~\cite{Huang_2020}, and a spin liquid phase~\cite{Wang_2018, Qian_2024}.

$\bullet$ In the checkerboard limit of $\delta=1$, the NNN hoppings on the $t_-$-plaquettes vanish.
Early studies on the anisotropic checkerboard lattice have primarily focused on using analytical methods to investigate metal-insulator transitions in the pyrochlore systems~\cite{Fujimoto_2002}, as well as the emergence of the plaquette-VBS phase at a finite
Hubbard interaction~\cite{Yoshioka_2008, Yoshioka_2008_1}.
For large $U$, the effective $J_1$-$J_2$ Heisenberg model provides a two-dimensional analog of the pyrochlore antiferromagnet~\cite{Starykh_2002, Singh_1998, Fouet_2003, Canals_2002}.
The existence of intermediate phases is still a topic of debate, with proposals such as the plaquette-VBS phase~\cite{Sindzingre_2002, Brenig_2004, Moukouri_2008, Chan_2011} and the possible N\'{e}el\textsuperscript{*} phase~\cite{Starykh_2005}, depending on the ratio of $J_2 / J_1$.
A deconfined phase transition has been proposed between the plaquette-VBS phase and the N\'{e}el phase~\cite{Senthil_2004}.
More recently, the effects of thermal fluctuations have also been taken into account using cellular dynamical MF theory~\cite{Liu_2019}.

In this work, we focus on an intermediate region $0 < \delta < 1$ to examine altermagnetic and stripe orders in both metallic and insulating phases, as well as their competition.
To be specific, we present our main results for $\delta=0.5$.
By comparing these results with those for other values of $\delta$, such as $\delta=0.2$ and $0.8$ shown in App.~\ref{app:appA}, we offer a general outline of the magnetism evolution driven by $\delta$.

\section{Mean-field analysis}
\label{sec:MeanfieldAnalysis}

In the simplest MF treatment, the on-site interaction is approximated as ${\hat n}_{l\uparrow}{\hat n}_{l\downarrow} \approx {\hat n}_{l\uparrow} \braket{\hat{n}_{l\downarrow}} + \braket{\hat{n}_{l\uparrow}} {\hat n}_{l\downarrow} - \braket{\hat{n}_{l\uparrow}} \braket{\hat{n}_{l\downarrow}}$.
This approximation assumes that one species of electrons move in the background of the site-dependent chemical potential provided by the other species.

The on-site superconducting term for the conventional $s$-wave pairing $\braket{\hat{c}^{\phantom{\dagger}}_{l\uparrow} \hat{c}^{\phantom{\dagger}}_{l\downarrow}}$, which breaks U($1$) symmetry, is neglected here because $U>0$, following the BCS paradigm~\cite{PhysRevB.36.857, PhysRevB.62.9083}.
We notice that spin-flipping terms $\braket{\hat{c}^{\dagger}_{l\uparrow} \hat{c}^{\phantom{\dagger}}_{l\downarrow}}$, describe gapless transverse spin excitations in the N\'{e}el phase.
In the stripe phase, these spin-flipping terms may play a role in fluctuating the orientations of spins along the stripe direction, which helps lower the energy on NN links in the spin-wave theory.
For simplicity, however, we omit them in the MF approximation.

Additionally, due to the absence of staggered on-site chemical potentials, NN repulsion, and 
other long-range interactions, a charge density wave is not present in the SU($2$) Hubbard model on the square lattice at half filling~\cite{Lin_1987}.
Thus we assume $n_{l\uparrow} + n_{l\downarrow} = 1$, with the definitions $n_{l\uparrow}=\braket{\hat{n}_{l\uparrow}}$ and $n_{l\downarrow}=\braket{\hat{n}_{l\downarrow}}$ in MF for any site $l$.

\subsection{Mean-field ansatz for magnetic structures}
\label{sec:MeanFieldAnsatzMagneticStructure}

For an arbitrary magnetic structure with finite magnetization $m$ and magnetic ordering wave vector $\mathbf{Q}$, we assume that $n_{l\uparrow/\downarrow} = 1/2 \pm m \exp(i \mathbf{Q} \cdot \mathbf{r}_l)$, where the vector $\mathbf{r}_l$ gives the position of site $l$~\cite{Hetzel_1998}.

In the MF scenario, we consider the magnetic structures of the N\'{e}el phase with a magnetic ordering wave vector $\mathbf{M} = (\pm\pi,\, \pm\pi)$, and the stripe AFM phase with a magnetic ordering wave vector, either $\mathbf{X} = (\pm\pi,\, 0)$ or $\mathbf{X}^\prime = (0,\, \pm\pi)$.
These two orders correspond to the ground states in the limits of $J_1 \gg J_2$ and $J_2 \gg J_1$ at large $U$, respectively.

For the N\'{e}el state, $n_{\text{A}\uparrow/\downarrow} = n_{\text{B}\downarrow/\uparrow} = 1/2 \pm m^{(\text{n})}$ are assumed for two inequivalent sublattice sites, where $m^{(\text{n})}$ denotes the \textit{N\'{e}el magnetization}.
This leads to the relation $2 m^{(\text{n})} = n_{\text{A}\uparrow} - n_{\text{A}\downarrow} = n_{\text{B}\downarrow} - n_{\text{B}\uparrow}$.
In the reciprocal space, the creation operators are defined as
\begin{eqnarray}
\left\{
\begin{array}{cc}
\hat{c}^\dag_{l\sigma} = \frac{1}{\sqrt{N}} \sum_{{\bf k} \in \text{FBZ}}\hat{a}^\dag_{{\bf k} \sigma} e^{-{\rm i}{\bf k}\cdot\mathbf{r}_l} & \quad \text{for} \quad l \in \text{A}\\\\
\hat{c}^\dag_{l\sigma} = \frac{1}{\sqrt{N}} \sum_{{\bf k} \in \text{FBZ}}\hat{b}^\dag_{{\bf k} \sigma} e^{-{\rm i}{\bf k}\cdot\mathbf{r}_l} & \quad \text{for} \quad l \in \text{B}
\end{array}
\right.
\end{eqnarray}
with $\hat{a}^\dag_{\mathbf{k}\sigma}$ and $\hat{b}^\dag_{\mathbf{k}\sigma}$ denoting the creation operators in reciprocal space.
The resulting MF Hamiltonian is given by
\begin{eqnarray}
\hat{H}^{(\text{n})} = \sum_{{\bf k} \in \text{FBZ}} \hat{\psi}^{(\text{n})\dag}_{{\bf k}} H^{(\text{n})}_{\bf k} \hat{\psi}^{(\text{n})}_{\bf k} + E^{(\text{n})}_0\, ,
\end{eqnarray}
where $\hat{\psi}^{(\text{n})\dag}_{\bf k} = ( \hat{a}^\dag_{{\bf k}\uparrow}$, $\hat{b}^\dag_{{\bf k}\uparrow}$, $\hat{a}^\dag_{{\bf k}\downarrow}$, $\hat{b}^{\dag}_{{\bf k}\downarrow} )$ is the spinor notation used as a basis for each lattice momentum ${\bf k}$~\cite{He_2024}.
The vacuum energy $E^{(\text{n})}_0 = N U [1 - 4 (m^{(\text{n})})^2]/2$ is given.
The $4 \times 4$ matrix $H^{(\text{n})}_{\bf k}$ is defined as
\begin{equation}
H^{(\text{n})}_{\bf k} = \left(
\begin{array}[c]{cccc}
\epsilon^+_{\bf k} + \mu_- & \epsilon^0_{\bf k} & & \\
\epsilon^0_{\bf k}   & \epsilon^-_{\bf k} + \mu_+ & & \\
 & & \epsilon^+_{\bf k} + \mu_+ & \epsilon^0_{\bf k} \\
 & & \epsilon^0_{\bf k} & \epsilon^-_{\bf k} + \mu_-
\end{array}
\right)
\end{equation}
with the NN structure factor $\epsilon^0_{\bf k} = -2 t (\cos k_x + \cos k_y)$, the NNN ones $\epsilon^\pm_{\bf k} = -2 t_\pm \cos(k_x + k_y) - 2 t_\mp \cos(k_x - k_y)$,
and $\mu_\pm = (1/2 \pm m^{(\text{n})}) U$.

For the stripe AFM state, we set $n_{\text{A}_1\uparrow/\downarrow} = n_{\text{B}_1\uparrow/\downarrow} = n_{\text{A}_2\downarrow/\uparrow} = n_{\text{B}_2\downarrow/\uparrow} = 1/2 \pm m^{(\text{s})}$, where $m^{(\text{s})}$ denotes the \textit{stripy magnetization}.
In the associated representation of $\hat{\psi}^{(\text{s})\dag}_{\bf k} = ( \hat{a}^\dag_{{\bf k}\uparrow}$, $\hat{b}^\dag_{{\bf k}\uparrow}$, $\hat{a}^\dag_{{\bf k} + \mathbf{X} \uparrow}$, $\hat{b}^\dag_{{\bf k} + \mathbf{X}\uparrow}$, $\hat{a}^\dag_{{\bf k}\downarrow}$, $\hat{b}^{\dag}_{{\bf k}\downarrow}$, $\hat{a}^\dag_{{\bf k} + \mathbf{X}\downarrow}$, $\hat{b}^{\dag}_{{\bf k} + \mathbf{X}\downarrow} )$, the MF Hamiltonian is written as
\begin{eqnarray}
\hat{H}^{(\text{s})} = \sum_{{\bf k} \in \text{MBZ}}
\hat{\psi}^{(\text{s})\dag}_{{\bf k}}
\begin{pmatrix}
H^{(\text{s})}_{\mathbf{k}\uparrow} & \\
& H^{(\text{s})}_{\mathbf{k}\downarrow}
\end{pmatrix}
\hat{\psi}^{(\text{s})}_{\bf k} + E^{(\text{s})}_0\, ,
\end{eqnarray}
with the vacuum energy $E^{(\text{s})}_0 = N U [1 - 4 (m^{(\text{s})})^2]/2$.
Notably, $\mathbf{k}$ sums over all modes in the MBZ, rather than the FBZ.
For given spin $\sigma$, the Hamiltonian component is
\begin{equation}
H^{(\text{s})}_{\mathbf{k}\sigma} = \left(
\begin{array}[c]{cccc}
\epsilon^+_{\bf k} + \mu_0 & \epsilon^0_{\bf k} & -s_\sigma \mu_1 & \\
\epsilon^0_{\bf k}   & \epsilon^-_{\bf k} + \mu_0 & & -s_\sigma\mu_1 \\
-s_\sigma \mu_1 & & \epsilon^+_{{\bf k} + {\bf X}} + \mu_0 & \epsilon^0_{{\bf k} + {\bf X}} \\
 & -s_\sigma\mu_1 & \epsilon^0_{{\bf k} + {\bf X}} & \epsilon^-_{{\bf k} + {\bf X}} + \mu_0
\end{array}
\right)\nonumber
\end{equation}
where $\mu_0 = U/2$ and $\mu_1 = m^{(\text{s})} U$.
The sign $s_\sigma=\pm 1$ corresponds to $\sigma=\uparrow$ and $\downarrow$, respectively.

In principle, a total of $4N$ variational parameters, specifically $n_{l\uparrow}$ and $n_{l\downarrow}$ for all sites, are necessarily determined.
In practice, considering translation symmetry, only two variational parameters, that is, magnetizations $m^{(\text{n})}$ and $m^{(\text{s})}$, are sufficient to capture the density imbalance of spins across distinct phases of the quantum phase diagram.
In the nonlinear variational process, we avoid the problem of metastable states by minimizing the energy starting from various initial guesses of the ground state.
More concretely, the initial magnetization values can be randomly chosen from the range $[0,\ 0.5]$.
However, since the exchanging symmetry between species may be broken, we restrict the initial states to satisfy $n_{\text{A}\uparrow} > n_{\text{A}\downarrow}$ and $n_{\text{B}\downarrow} > n_{\text{B}\uparrow}$.

\subsection{Phases and phase transitions}
\label{sec:MeanFieldPhasePhaseTransition}

\begin{table}[b!]
\begin{threeparttable}[b]
    \centering
    \begin{tabular}{|c|c|c|c|c|c|c|c|}
	\hline
        \multirow{3}*{Transition} & \multicolumn{7}{|c|}{Order} \\
        \cline{2-8}
                                               & \multicolumn{4}{|c|}{MF} & \multicolumn{3}{|c|}{DMRG} \\
        \cline{2-8}
	                                      & TDL & $8$A & $16$B & $24$B & $8$A & $16$B & $24$B \\
	\hline
    	NM\textsubscript{1}-NM\textsubscript{2} & Lifshitz & $\slash$ & $\slash$ & $\slash$ & $\slash$ & $\slash$ & $\slash$ \\
	\hline
	NM\textsubscript{2}-NM\textsubscript{3} & Lifshitz & $\slash$ & $\slash$ & $\slash$ & $\slash$ & $\slash$ & $\slash$ \\
	\hline
	SM\textsubscript{1}-SM\textsubscript{2} & Lifshitz & $\slash$ & $\slash$ & $\slash$ & $\slash$ & $\slash$ & $\slash$ \\
	\hline
	AMM\textsubscript{1}-AMM\textsubscript{2} & Lifshitz & $\slash$ & $\slash$ & $\slash$ & $\slash$ & $\slash$ & $\slash$ \\
	\hline
	NM\textsubscript{2}-SM\textsubscript{2} & 1st & $\slash$ & $\slash$ & $\slash$ & $\slash$ & $\slash$ & $\slash$ \\
         \hline
         NM\textsubscript{1}-AMM\textsubscript{1} & 1st & $\slash$ & $\slash$ & $\slash$ & $\slash$ & $\slash$ & $\slash$ \\
         \hline
         NM\textsubscript{2}-AMM\textsubscript{2} & 1st & $\slash$ & $\slash$ & $\slash$ & $\slash$ & $\slash$ & $\slash$ \\
         \hline
         AMM\textsubscript{1}-SM\textsubscript{1} & 1st & $\slash$ & 1st & 1st & $\slash$ & 1st & 1st \\
	\hline
         AMM\textsubscript{1}-SM\textsubscript{2} & 1st & $\slash$ & $\slash$ & $\slash$ & $\slash$ & $\slash$ & $\slash$ \\
	\hline
	AMM\textsubscript{2}-SM\textsubscript{2} & 1st & $\slash$ & $\slash$ & $\slash$ & $\slash$ & $\slash$ & $\slash$ \\
	\hline
         AMM\textsubscript{1}-AMI & 2nd & 1st & 1st & 1st & 1st & 1st & 1st \\
	\hline
	AMI-SI & 1st & 1st & 1st & 1st & 1st & $\times$ & $\times$ \\
	\hline
        NM\textsubscript{3}-SM\textsubscript{1} & Unknown\tablefootnote & $\slash$ & 2nd & 2nd & $\slash$ & $\times$ & 2nd \\
         \hline
        SM\textsubscript{1}-SI & 2nd & $\slash$ & 1st & 1st & $\slash$ & $\times$ & $\times$ \\
         \hline
        AMM\textsubscript{1}-SI & $\slash$ & 1st & 1st & 1st & 1st & $\slash$ & $\slash$ \\
         \hline
        AMM\textsubscript{1}-NM\textsubscript{3} & $\slash$ & 1st & 1st & 1st & $\slash$ & 1st & 1st \\
         \hline
          NM\textsubscript{3}-SI & $\slash$ & 2nd & $\slash$ & $\slash$ & 2nd\tablefootnote & $\slash$ & $\slash$ \\
         \hline
         AMM\textsubscript{1}-VBS & $\slash$ & $\slash$ & $\slash$ & $\slash$ & $\slash$ & $\slash$ & 1st \\
         \hline
         VBS-SI & $\slash$ & $\slash$ & $\slash$ & $\slash$ & $\slash$ & $\slash$ & 1st \\
         \hline
    \end{tabular}
    \begin{tablenotes}
    \item [a] The magnetization $m^{(\text{s})}$ increases very slowly from zero, making it difficult to pinpoint the exact order of the transition.
    \item [b] A crossover occurs when $t' \gtrsim 0.8$.
     \end{tablenotes}
    \caption{The order of transitions summarized from the MF and DMRG-calculated ground-state phase diagrams in Figs.~\ref{fig:fig3}(c1-c3) and (d1-d3) for $8$A, $16$B, and $24$B tori, as well as in Fig.~\ref{fig:fig1}(d) in the TDL.
    The notation ``$\slash$" indicates that the transition is absent for the specified case, while the notation ``$\times$" signifies that the transition is replaced by a crossover.}
    \label{tab:tabel1}
\end{threeparttable}
\end{table}

\begin{figure*}[t!]
\centering
\includegraphics[width=0.9\linewidth]{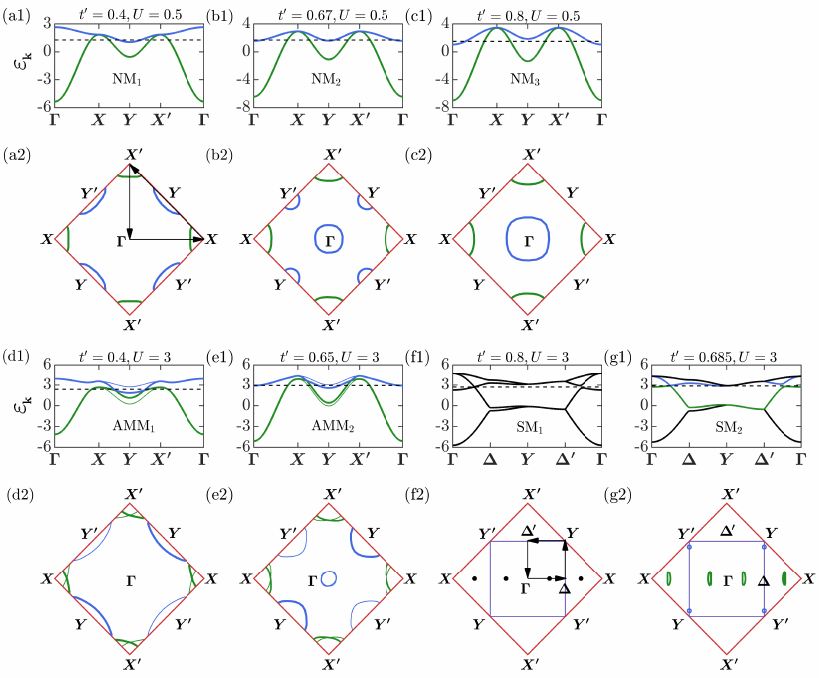}
\caption{(a1-g1) The energy dispersions, and (a2-g2) the corresponding hole-type (green) and electron-type (blue) Fermi surfaces in the FBZ, are displayed for various parameter sets in (a1,a2) NM\textsubscript{1}, (b1,b2) NM\textsubscript{2}, (c1,c2) NM\textsubscript{3}, (d1,d2) AMM\textsubscript{1}, (e1,e2) AMM\textsubscript{2}, (f1,f2) SM\textsubscript{1}, and (g1,g2) SM\textsubscript{2}.
In (a1-e1), the energy dispersions are shown along a black path highlighted by a sequence of arrows in (a2), while in (f1,g1), a different path is chosen, as marked in (f2).
The black dashed lines in the energy dispersions represent the Fermi energy at half filling.
We present the spin-resolved data, where thick and thin solid lines correspond to spin-up and spin-down species, respectively.
}
\label{fig:fig2}
\end{figure*}

In the TDL, as shown in Fig.~\ref{fig:fig1}(d), we identify nine distinct phases: three NM phases, two AMM phases, an AMI phase, two SM phases, and an SI phase.
At cutting lines $U=0.5$, $3$ and $7$, the variational magnetizations $m^{(\text{n})}$ and $m^{(\text{s})}$ are plotted in Figs.~\ref{fig:fig1}(e1-e3).
By analyzing the first/second-order derivatives of the ground-state energy and the charge gap, as presented in Figs.~\ref{fig:fig1}(f1-f3), we can determine the order of various phase transitions, which are summarized in Table.~\ref{tab:tabel1}.
The charge gap
\begin{eqnarray}
\Delta_\text{c} ({\bf k}) = \min_{{\bf k}_1, {\bf k}_2}[E_2({\bf k}_2) - E_1({\bf k}_1)]\, \label{eq:chargegap}
\end{eqnarray}
is defined as the minimum energy difference between the first unoccupied band $E_2$ at momentum $\mathbf{k}_2$ and the last occupied band $E_1$ at momentum $\mathbf{k}_1$.

The collinear magnetic structures of the N\'{e}el and stripe AFM states correspond to different point-group symmetries.
For an aspect ratio of $L_x / L_y = 1$, the N\'{e}el phase is invariant under $\pi/2$-rotations about a lattice site, while the stripe AFM phase is not.
Therefore, the transitions that involve a change in magnetic structure from the N\'{e}el order to the stripe AFM order are of first order, remarkably in line with the Landau-Ginzburg-Wilson theory~\cite{Senthil_2004}.
For example, as $t^\prime$ is increased along the cutting line $U=3$, we observe that the magnetization for the N\'{e}el phase, $m^{(\text{n})}$, suddenly drops to zero at the AMM\textsubscript{2}-SM\textsubscript{2} transition point $t^\prime \approx 0.6716$, as shown in Fig.~\ref{fig:fig1}(e2).
This drop is accompanied by the emergence of nonzero magnetization for the stripe AFM phase, $m^{(\text{s})}$.
At this transition point, both the first-order derivative (not shown) and the second-order derivative of the ground-state energy, as shown in Fig.~\ref{fig:fig1}(f2), exhibit clear singularities, confirming the first-order nature of the transition.
At the AMM\textsubscript{1}-AMM\textsubscript{2} and SM\textsubscript{1}-SM\textsubscript{2} transition points, we observe a jump and a kink, respectively, in the second-order derivative of the ground-state energy, as shown in Fig.~\ref{fig:fig1}(f2).
Both are the Lifshitz-type transitions, associated with changes in the Fermi surfaces, as discussed later in Figs.~\ref{fig:fig2}(d-g).

Similarly, in Figs.~\ref{fig:fig1}(e3,f3), along another cutting line $U=7$, we find two jumps in both $m^{(\text{n})}$ and $m^{(\text{s})}$ at the AMI-SI transition point $t^\prime\approx 0.6529$, without a closure of the charge gap $\Delta_\text{c}$.
This supports the idea that this transition is first-order.
Nevertheless, the Hamiltonian~\eqref{eq:model} lacks $\pi/2$-rotations about a lattice site due to $\delta \ne 0$, which may lead to modifications in the N\'{e}el phase that break this symmetry after considering quantum fluctuations.
In that case, we might observe a crossover instead, as shown later in Figs.~\ref{fig:fig6}(f2-f3).
In contrast, if this symmetry is restored in the altermagnetic orders for a given torus geometry, the transition could become first order, as shown later in Figs.~\ref{fig:fig6}(e1-e3).

\begin{figure*}[t!]
\centering
\includegraphics[width=0.95\linewidth]{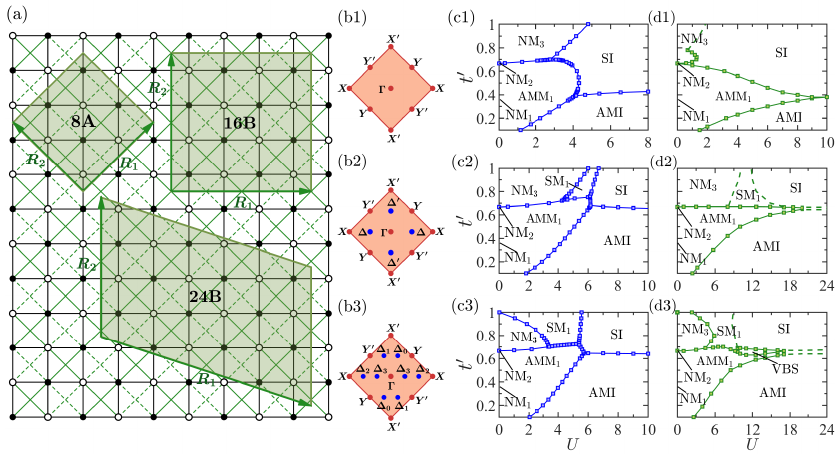}
\caption{
(a) A parallelogram (green-shaded) is sheared on the square lattice and then rolled into a torus along the vectors $\mathbf{R}_1$ and $\mathbf{R}_2$ (green arrows).
The allowed momentum points in the FBZ are marked for the (b1) $8$A, (b2) $16$B and (b3) $24$B tori.
Notably, all high-symmetry points $\mathbf{\Gamma}$, $\mathbf{X}$, $\mathbf{X}^\prime$, $\mathbf{Y}$ and $\mathbf{Y}^\prime$ are selected in these tori and highlighted with red dots.
In (b2,b3), other momentum points $\mathbf{\Delta} = (\pm\pi/2,\, 0)$, $\mathbf{\Delta}^\prime = (0,\, \pm\pi/2)$, $\mathbf{\Delta}_0 = (\pm\pi/6,\, \pm\pi/2)$, $\mathbf{\Delta}_1 = (\mp\pi/6,\, \pm\pi/2)$, $\mathbf{\Delta}_2 = (\pm2\pi/3,\, 0)$ and $\mathbf{\Delta}_3 = (\pm\pi/3,\, 0)$ are colored in blue.
For these geometries, (c1-c3) the MF and (d1-d3) DMRG-calculated ground-state phase diagrams are given, respectively.
The phase boundaries are highlighted by (c1-c3) blue and (d1-d3) green lines.
}
\label{fig:fig3}
\end{figure*}

\subsection{Fermi surface topology}
\label{sec:MeanFieldFermiSurfaceTopology}

To characterize the different metallic phases, we examine the topology of the Fermi surfaces.
For the three NM phases and two AMM phases shown in Figs.~\ref{fig:fig2}(a-e), the electron-type Fermi surfaces are centered at either two inequivalent $Y$-points, $\mathbf{Y} = (\pm\pi/2,\, \pm\pi/2)$ and $\mathbf{Y}^\prime = (\pm\pi/2,\, \mp\pi/2)$, or at $\mathbf{\Gamma} = (0,\, 0)$, or at both.
Meanwhile, the hole-type Fermi surfaces are centered at two inequivalent $X$-points, i.e., $\mathbf{X}$ and $\mathbf{X}^\prime$.
As shown in Figs.~\ref{fig:fig2}(d,e), in the AMM\textsubscript{1} and AMM\textsubscript{2} phases, the hole-type Fermi surfaces near $\mathbf{X}$ and $\mathbf{X}^\prime$, as well as the electron-type Fermi surfaces near $\mathbf{\Gamma}$, display a mismatch between the spin-up and spin-down electrons.
Additionally, the electron-type Fermi surfaces near $\mathbf{Y}$ and $\mathbf{Y}'$ are constructed by the spin-up and spin-down electrons, respectively.
These features clearly indicate altermagnetic behavior in the AMM\textsubscript{1} and AMM\textsubscript{2} phases~\cite{Reimers_2024, PhysRevB.110.205120, PhysRevB.110.205140, rajpurohit_2024, durrnagel_2024, chen_2024, kaushal_2024}.

In the SM\textsubscript{1} phase, as shown in Fig.~\ref{fig:fig2}(f), the Fermi surfaces are replaced with Dirac points, sitting on the $k_x$-axis.
It is worth noting that, due to the combination of time-reversal and translation operations being equivalent to a $\pi$-rotation about a lattice site in the stripe AFM phase, the energy spectrum for the two spin species remains degenerate at all momenta~\cite{Cui_2023}.
In the MBZ, the positions of the two inequivalent Dirac points can be tuned by adjusting the hopping amplitude $t^\prime$.
In the special case where $t^\prime=1$, the Dirac points are located at $(\pm 2\pi/3,\, 0)$ and $(\pm \pi/3,\, 0)$, that is, $\mathbf{\Delta}_2$ and $\mathbf{\Delta}_3$ as marked in Fig.~\ref{fig:fig3}(b3).
In another stripe metallic phase SM\textsubscript{2}, as shown in Fig.~\ref{fig:fig2}(g), these Dirac points evolve into hole-type petaloid Fermi surfaces, while electron-type circular Fermi surfaces emerge near $\mathbf{Y}$ and $\mathbf{Y}^\prime$, thereby restoring the half-filling condition.

At the NM\textsubscript{1}-NM\textsubscript{2} transition, the appearance of an electron-type Fermi surface centered at $\mathbf{\Gamma}$ signals a Lifshitz transition, consistent with the continuous nature of the NM\textsubscript{1}-NM\textsubscript{2} transition, as listed in Table~\ref{tab:tabel1}.
Similarly, the NM\textsubscript{2}-NM\textsubscript{3}, AMM\textsubscript{1}-AMM\textsubscript{2}, and SM\textsubscript{1}-SM\textsubscript{2} transitions are also of the Lifshitz type~\cite{Lifshitz_1960}.

From the above MF analysis, we find that the lowest-energy excitations at the points $\mathbf{\Gamma}$, $\mathbf{X}$, $\mathbf{X}^\prime$, $\mathbf{Y}$ and $\mathbf{Y}^\prime$ are crucial for understanding the key features of the phases and phase transitions.
Therefore, in the next section on the DMRG calculations, we choose appropriate torus geometries to guarantee that the relevant low-energy excitations hit these high-symmetry momentum points~\cite{He_2017, Hu_2019, He_2024}.

Here, we want to emphasize that the MF analysis does not account for quantum fluctuations sufficiently, particularly in the intermediate-$t^\prime$ region.
Strong quantum fluctuations could potentially lead to the emergence of unconventional quantum phases.
In the next section, our DMRG results indicate that the microscopic physical scenario may qualitatively change in the altermagnetic and stripe phases.

\section{DMRG calculations}\label{sec:DMRGCalculations}

\subsection{Torus geometry selection}\label{sec:TorusGoemetrySelection}
In DMRG calculations, the first step is to choose a finite-size cluster.
As mentioned in the previous section, the geometry of the cluster should be carefully selected so that relevant high-symmetry momentum points in the FBZ can be hit.
To begin, we shear a parallelogram with two sides denoted by the vectors $\mathbf{R}_1$ and $\mathbf{R}_2$, respectively.
For example, in the case of the $8$A torus~\cite{Maier_2005}, we set $\mathbf{R}_1 = (2,\, 2)$ and $\mathbf{R}_2 = (-2,\, 2)$, as shown in Fig.~\ref{fig:fig3}(a).
When the geometry is rolled up into a torus, the Hamiltonian~\eqref{eq:model} remains invariant under translations by either $\mathbf{R}_1$ or $\mathbf{R}_2$.
This implies that the allowed momentum $\mathbf{k}$ must satisfy the following conditions
\begin{eqnarray}\label{eq:selectedmomentum}
\mathbf{k} \cdot \mathbf{R}_1 = 2 \pi m\, ,\quad \mathbf{k} \cdot \mathbf{R}_2 = 2 \pi n\, ,
\end{eqnarray}
where $m$ and $n$ are arbitrary integers.
As the number of lattice sites $2N$ increases, 
the geometry must be carefully selected to maintain the ability to hit high-symmetry momentum points.
Due to computational limitations, we mainly consider the $8$A, $16$B and $24$B tori.
The allowed momentum points for these geometries, as indicated in Fig.~\ref{fig:fig3}(b1-b3), cover all the relevant high-symmetry points~\cite{He_2017, Hu_2019, He_2024}.
In the following subsections, we will systematically compare the phase diagrams obtained from the DMRG calculations and the MF analysis.

For small systems, the following apparent observations in the ground-state phase diagram, as shown in Figs.~\ref{fig:fig3}(c1-c3,d1-d3), are worth noting: (1) Both the NM\textsubscript{2} and AMM\textsubscript{2} phase regions are absent because the momentum points associated with the very narrow Fermi surface emerging near $\mathbf{\Gamma}$ are not hit. In particular, the NM\textsubscript{2} phase survives only at the transition points between the NM\textsubscript{1} and NM\textsubscript{3} phases.
(2) The SM\textsubscript{2} phase is also absent, as only a single momentum point near $\mathbf{\Delta}_2$ is likely to be hit, resembling the unique ``Dirac point" in SM\textsubscript{1}. In this case, the SM\textsubscript{2} phase is indistinguishable from the SM\textsubscript{1} one and, in principle, can be attributed to the latter. (3) NM\textsubscript{1} survives only at $U=0$, as even a tiny $U$ is sufficient to induce altermagnetism when $t^\prime < 2/3$.

Unless stated explicitly otherwise, we use truncated bond dimensions of $\chi = 4,000$ for the $8$A and $16$B tori, and $\chi=6,000$ for the $24$B torus in the U($1$)$\times$SU($2$) DMRG algorithm.
Additionally, we perform at least $14$ sweeps during the variational process to ensure that the truncation errors are on the order of $10^{-5}$.

\subsection{Magnetic metallic and insulating phases}
\label{sec:MagneticMetallicInsulating}
In the altermagnetic phases, the ground state, i.e., the N\'{e}el state, exhibits spin rotation symmetry breaking in the TDL, and the low-energy excitation spectrum contains gapless Goldstone modes~\cite{Goldstone1962, Low2002}.
However, for these small tori with limited system sizes, the spin rotation symmetry in the ground states of the AMM\textsubscript{1} and AMI phases remains unbroken.
As a result, detecting energy spectrum splits or mismatches for spin species, as suggested by the MF scenario in Fig.~\ref{fig:fig2}(d), becomes challenging in DMRG simulations.
Instead, we calculate the spin correlations
\begin{eqnarray}
G_{l l^\prime} = \braket{ \hat{\mathbf{S}}_l \cdot  \hat{\mathbf{S}}_{l^\prime} }\, ,\quad
\end{eqnarray}
and the corresponding spin structure factor
\begin{eqnarray}
S(\mathbf{k}) = \frac{1}{4 N^2} \sum_{l,l^{\prime}} e^{i \mathbf{k} \cdot (\mathbf{r}_l - \mathbf{r}_{l^\prime})} G_{l l^\prime}\, .\quad
\end{eqnarray}
to determine the relevant phases.
Here, the spin vector operator at site $l$ is defined as
\begin{eqnarray}
\hat{\mathbf{S}}_l = \sum_{\sigma, \sigma^\prime} \hat{c}_{l \sigma}^\dag (S^x,\, S^y,\, S^z)_{\sigma\sigma^\prime} \hat{c}^{\phantom{\dag}}_{l \sigma^\prime}\, ,
\end{eqnarray}
where $S^x$, $S^y$ and $S^z$ give the matrix representations for the spin-half operators.
Additionally, we investigate the momentum distribution $n(\mathbf{k})$ to examine the occupation in momentum space, which is given by
\begin{eqnarray}
n(\mathbf{k}) = \frac{1}{N} {\overline{\sum}}_{l,l^\prime} \sum_\sigma e^{i \mathbf{k} \cdot (\mathbf{r}_l - \mathbf{r}_{l^\prime})} \braket{\hat{c}^\dag_{l\sigma} \hat{c}^{\phantom{\dag}}_{l^\prime \sigma}}\, ,
\end{eqnarray}
where ${\overline{\sum}}_{l,l^\prime}$ denotes the summation over pairs of sites $l$ and $l^\prime$ that both belong to either sublattices A or B.

\begin{figure*}[t!]
\centering
\includegraphics[width=0.95\linewidth]{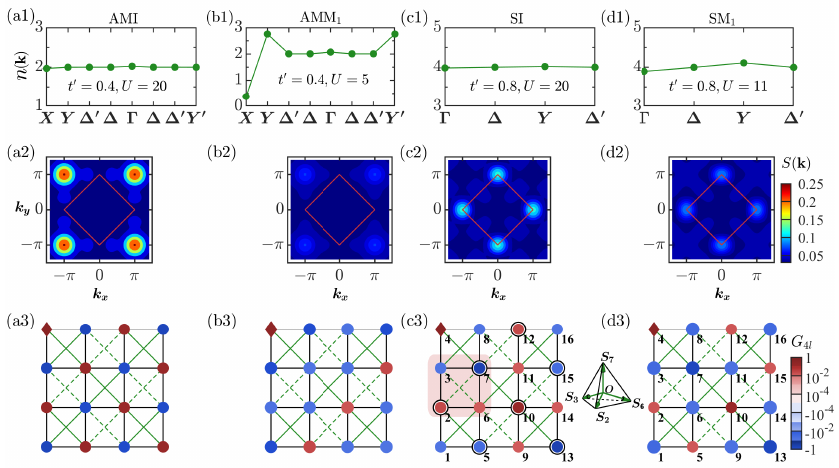}
\caption{(a1-d1) The momentum distributions $n(\mathbf{k})$ measured along the paths marked in Figs.~\ref{fig:fig2}(a2,f2), (a2-d2) the spin structure factors $S(\mathbf{k})$ in the FBZ, and (a3-d3) the spin correlations $G_{4 l}$ starting from the target site $4$ (\tcr{$\sblacklozenge[0.8]$}), for (a1-a3) AMI, (b1-b3) AMM\textsubscript{1}, (c1-c3) SI, and (d1-d3) SM\textsubscript{1} in the $16$B torus.
In (a3-d3), the magnitude of spin correlations are indicated by colored circles. (see text)
The spin orientations at sites $2$, $3$, $6$ and $7$ within a unit cell (shaded in red) form a regular tetrahedron.
}
\label{fig:fig4}
\end{figure*}

In Fig.~\ref{fig:fig4}(a1), for the AMI phase, all allowed momentum points are uniformly occupied by two electrons with distinct spin species.
The upper and lower bands are separated by a finite gap, which is a characteristic property of an insulator.
In Fig.~\ref{fig:fig4}(a2), the corresponding spin structure factor shows characteristic peaks at $\mathbf{M}$, with the staggered magnetization defined as $M^{(\text{n})} = \sqrt{S(\mathbf{M})}$.
These peaks are consistent with the alternating spin correlations along both the $x$ and $y$-axes, observed in Fig.~\ref{fig:fig4}(a3).

Differently, in Fig.~\ref{fig:fig4}(b1), the AMM\textsubscript{1} phase has nearly one additional electron occupying the high-symmetry points $\mathbf{Y}$ and $\mathbf{Y}^\prime$, consistent with the formation of electron-type Fermi surfaces centered at these points, as suggested by MF and shown in Fig.~\ref{fig:fig2}(d2).
At half filling, this balances the hole filling at $\mathbf{X}$, which is surrounded by hole-type Fermi surfaces.
Therefore, we can find nearly-empty occupation at $\mathbf{X}$ in Fig.~\ref{fig:fig4}(b1).
Compared to the AMI phase, the AMM\textsubscript{1} phase exhibits similar features in the spin structure factor, with relatively lower peaks at $\mathbf{M}$, as seen in Fig.~\ref{fig:fig4}(b2).
However, due to the mobility of electrons, an electron prefers to be surrounded by holes, or a cloud of electrons with reversed spin, which helps lower the energy in AMM\textsubscript{1}.
As a result, Fig.~\ref{fig:fig4}(b3) shows that the characteristics of staggered spin correlations starting from the target site $4$, are screened and can not be observed at short distances, but are expected to be restored at longer distances in a larger cluster.

For the stripe orders in the SI and SM\textsubscript{1} phases, we define the stripy magnetization as $M^{(\text{s})} = \sqrt{\max(S(\mathbf{X}), S(\mathbf{X^\prime}))}$, where the spin structure factors at the points $\mathbf{X}$ and $\mathbf{X}^\prime$ may differ.
Since the two equivalent Dirac points, indicated in Fig.~\ref{fig:fig2}(f2), do not touch within the $16$B torus, both the SI and SM\textsubscript{1} phases have finite charge gaps.
The allowed momentum points $\mathbf{\Gamma}$, $\mathbf{\Delta}$, $\mathbf{\Delta}^\prime$ and $\mathbf{Y}$ are uniformly occupied by four electrons in the MBZ: two spin-up and two spin-down electrons, as shown in Figs.~\ref{fig:fig4}(c1,d1).
In Figs.~\ref{fig:fig4}(c2,d2), the spin structure factors for these stripe phases exhibit peaks at the $\mathbf{X}$ and $\mathbf{X}^\prime$ points, which are typically indicative of stripe AFM ordering along either the $x$ or $y$-axis, as suggested by MF.
These peaks do not collapse into either $\mathbf{X}$ or $\mathbf{X}^\prime$ due to the absence of spontaneous breaking of the $\pi/2$-rotational symmetry for small lattice sizes.

Quantum fluctuations play a significant role in the SI phase, causing the spins to deviate from strict collinearity and orientate out of the plane, especially when the effective spin couplings for the strong and weak NNN links chosen in this work satisfy $J_+ / J_- = \left(t_+/t_-\right)^2 = 9 \gg 1$.
Following the classical variational for spins~\cite{Chen_2023}, a pair of spins along the diagonal of the $t_+$-plaquettes remain exactly antiparallel in the ground state.
Meanwhile, the spins on the $t_-$-plaquettes, such as those at sites $2$, $3$, $6$ and $7$ within a unit cell marked in Fig.~\ref{fig:fig4}(c3) and the inset, orientate in four distinct directions: two spins stagger in a plane, while the other two are tilted out of the plane, creating a regular tetrahedron.
In four-fold degenerate ground-state manifolds, the target spin at site $4$ consistently aligns parallel to the spins at sites $2$, $10$ and $12$, while keeping antiparallel to the spins at sites $5$, $7$, $13$ and $15$.
These findings are verified by the DMRG-calculated spin correlations between the target spin and the sites marked with black circles in Fig.~\ref{fig:fig4}(c3).
In contrast, the spin correlations at other sites are relatively small.
Similarly, in the SM\textsubscript{1} phase, the motion of electrons slightly violates the spin order at short distances, such as the spin correlation between sites $4$ and $15$, as shown in Fig.~\ref{fig:fig4}(d3).
This disruption also results in a reduction of the peak intensity of the spin structure factors at $\mathbf{X}$ and $\mathbf{X}^\prime$, as depicted in Fig.~\ref{fig:fig4}(d2).

\begin{figure}[t!]
\centering
\includegraphics[width=\linewidth]{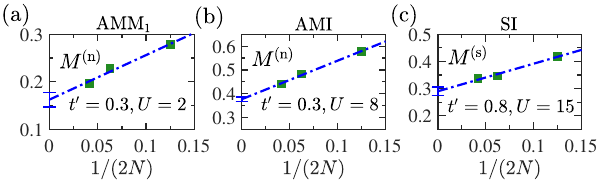}
\caption{The linear extrapolations of the magnetizations (\tcg{$\sblacksquare[0.7]$}) $M^{(\text{n})}$ and $M^{(\text{s})}$ as $N$ approaches infinity.
In the TDL, the magnetizations are as follows: (a) $M^{(\text{n})} \approx 0.1627(15)$ for AMM\textsubscript{1}, (b) $M^{(\text{n})} \approx 0.377(9)$ for AMI, and (c) $M^{(\text{s})} \approx 0.290(15)$ for SI.
}
\label{fig:fig5}
\end{figure}

\begin{figure*}[t!]
\centering
\includegraphics[width=0.95\linewidth]{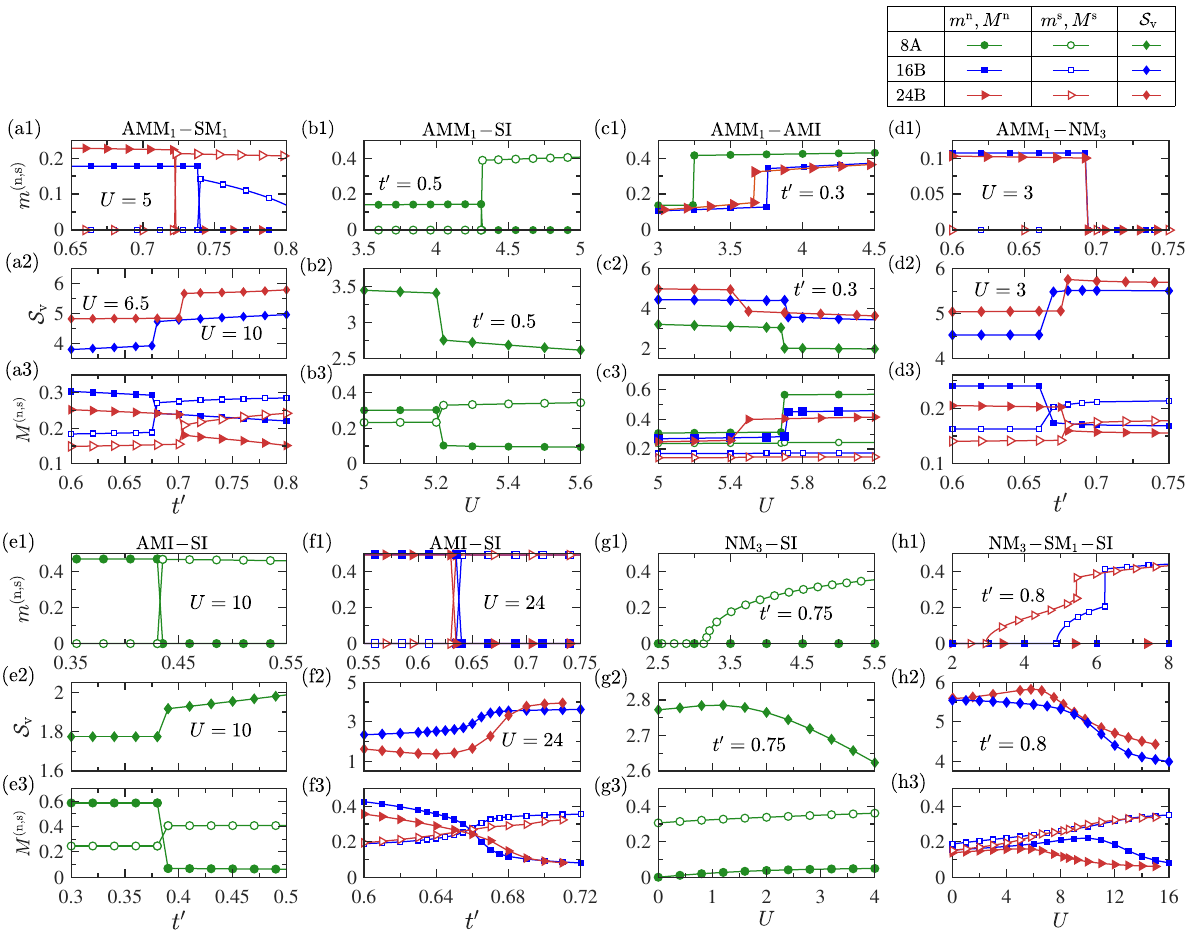}
\caption{The magnetizations $m^{(\text{n})}$ (\tcg{$\sbullet[1.1]$}, \tcb{$\sblacksquare[0.7]$}, \tcr{$\sblacktriangleright[0.9]$}) and $m^{(\text{s})}$ (\tcg{$\scirc[1.1]$}, \tcb{$\sBox[0.7]$}, \tcr{$\svartriangleright[0.9]$}) in the MF analysis, the DMRG-calculated von Neumann entanglement entropy $\mathcal{S}_\text{v}$ (\tcg{$\sblacklozenge[0.8]$}, \tcb{$\sblacklozenge[0.8]$}, \tcr{$\sblacklozenge[0.8]$}) and magnetizations $M^{(\text{n})}$ (\tcg{$\sbullet[1.1]$}, \tcb{$\sblacksquare[0.7]$}, \tcr{$\sblacktriangleright[0.9]$}) and $M^{(\text{s})}$ (\tcg{$\scirc[1.1]$}, \tcb{$\sBox[0.7]$}, \tcr{$\svartriangleright[0.9]$}), as functions of either the hopping amplitude $t^\prime$, or the on-site repulsion strength $U$, for the $8$A (green), $16$B (blue), $24$B (red) tori.
In the upper panels, we concern four distinct first-order phase transitions: (a1-a3) AMM\textsubscript{1}-SM\textsubscript{1},
(b1-b3) AMM\textsubscript{1}-SI,
(c1-c3) AMM\textsubscript{1}-AMI,
and (d1-d3) AMM\textsubscript{1}-NM\textsubscript{3}.
In the lower panels, we focus on the phase transitions and crossovers: (e1-e3,f1-f3) AMI-SI, (g1-g3) NM\textsubscript{3}-SI, and (h1-h3) NM\textsubscript{3}-SM\textsubscript{1}-SI.
}
\label{fig:fig6}
\end{figure*}

In Fig.~\ref{fig:fig5}, we plot the magnetizations $M^{(\text{n})}$ and $M^{(\text{s})}$ as functions of $1/(2N)$ for three different tori.
For typical parameters in AMM\textsubscript{1}, AMI, and SI, the linear extrapolations of the magnetizations to the TDL suggest finite values within errorbars, which serves a hallmark of the existence of magnetic orders.
Notably, the peak heights of spin structure factors at $\mathbf{X}$ and $\mathbf{X}^\prime$ are the same for the $8$A and $16$B tori, while the maximum of the spin structure factor for the $24$B torus is only located at $\mathbf{X}$ due to the absence of $\pi/2$-rotational symmetry with respect to the center of an arbitrary plaquette in this geometry.
Here, we cannot do the extrapolation for the SM\textsubscript{1} phase, since this phase is missing for the $8$A torus, as shown in Fig.~\ref{fig:fig3}(d1).

\subsection{First order transitions}\label{FirstOrderTransition}

In the following analysis, we compare the ground-state phase diagrams obtained from the MF analysis and the DMRG calculations to gain a better understanding of the physical scenario for each torus geometry.

We use the bipartite von Neumann entanglement entropy
\begin{eqnarray}\label{eq:entropy}
\mathcal{S}_\text{v} = -\text{tr} (\rho^{(\triangleleft)} \ln \rho^{(\triangleleft)})
\end{eqnarray}
to determine the transition points in the ground-state phase diagram~\cite{He_2024_1}.
Here, the reduced density matrix for the left semichain $\triangleleft$ is given by $\rho^{(\triangleleft)} = \text{tr}_\triangleright \ketbra{\psi}{\psi}$, after tracing out degrees of freedom for the right semichain $\triangleright$ from the ground-state wave function $\ket{\psi}$.
In Fig.~\ref{fig:fig6}, we concern four first-order phase transitions in the ground-state phase diagrams shown in Figs.~\ref{fig:fig3}(c1-c3) and (d1-d3).
These transitions are expected to appear as discontinuities in the curves of both von Neumann entanglement entropy and the magnetizations.
Using this criterion, we can easily identify the positions of transition points for the AMM\textsubscript{1}-SM\textsubscript{1} and AMM\textsubscript{1}-SI transitions.

We use the AMM\textsubscript{1}-SM\textsubscript{1} transition as an example.
In Fig.~\ref{fig:fig6}(a1), along the cutting line $U=5$, the curves of the staggered magnetization $m^{(\text{n})}$ in the MF analysis exhibit jumps at $t^\prime \approx 0.7382$ and $0.7223$ for the $16$B and $24$B tori, respectively.
At these points, the stripy magnetization $m^{(\text{s})}$ also emerges from zero suddenly.
In comparison, along the cutting line $U=10$ for $16$B and another cutting line $U=6.5$ for $24$B, the DMRG-calculated entanglement entropy $\mathcal{S}_\text{v}$ and staggered magnetization $M^{(\text{n})}$ both show jumps at different locations $t^\prime \approx 0.675$ and $0.7$, repsectively, as shown in Figs.~\ref{fig:fig6}(a2,a3).
Similarly, Figs.~\ref{fig:fig6}(b1-b3) indicate the discontinuities for the AMM\textsubscript{1}-SI transition, and we summarize their transition order in Table~\ref{tab:tabel1}.

In Figs.~\ref{fig:fig6}(c1-c3), we also find clear jumps at the AMM\textsubscript{1}-AMI transition for all three tori.
As shown in Fig.~\ref{fig:fig6}(c1), the staggered magnetization $m^{(\text{n})}$ curves in MF exhibit jumps at $U \approx 3.24$, $3.75$ and $3.66$ for the $8$A, $16$B and $24$B tori, respectively.
Correspondingly, the DMRG-calculated entanglement entropy $\mathcal{S}_\text{v}$ and staggered magnetization $M^{(\text{n})}$ in Figs.~\ref{fig:fig6}(c2,c3) also show jumps at different places $U\approx 5.68$, $5.7$ and $5.4$.
The heights of these jumps decrease gradually as the torus geometry changes from $8$A to $24$B, suggesting that these first order discontinuities are influenced by finite-size effects.
In the TDL, the MF analysis suggests that the AMM\textsubscript{1}-AMI transition becomes continuous, as indicated by the smooth variation of $m^\text{(\text{n})}$ near $t^\prime \approx 0.267$ at $U=3$ in Fig.~\ref{fig:fig1}(e2).
Thus, we expect that the discontinuities of $\mathcal{S}_\text{v}$ and $M^{(\text{n})}$ in the DMRG calculations may also disappear in the TDL.

Similarly, the AMM\textsubscript{1}-NM\textsubscript{3} transition, shown in Figs.~\ref{fig:fig6}(d1-d3), is also of first order.
Additionally, for other torus geometries, as discussed in App.~\ref{app:appB}, where not all high-symmetry points are hit, we do not observe discontinuities in the curves.

Lastly, we investigate the direct transition from the AMI phase to the SI phase at dominantly-large $U$.
For the $8$A torus, the ground-state of the AMI phase restores the $\pi/2$-rotational symmetry with respect to an arbitrary site (not shown).
As a result, we observe discontinuities in the curves of the DMRG-calculated entanglement entropy and magnetizations in the vicinity of the transition to the SI phase, as shown in Figs.~\ref{fig:fig6}(e2,e3), consistent with the MF results presented in the panel (e1).
Nevertheless, for the $16$B and $24$B tori, the transition undergoes two successive processes: (\textbf{1}) Spins along the diagonal of the $t_+$-plaquettes align antiparallel to lower the energy of the $t_+$ NNN links. (\textbf{2}) The four distinct spin orientations in the $t_-$-plaquettes adjust to further reduce the energy of the $t_-$ NNN links.
In these processes, the energy of the horizontal and vertical NN links changes passively because of $t_+ \gg t_-$.
As a result, the DMRG-calculated entanglement entropy and magnetizations, shown in Figs.~\ref{fig:fig6}(f2,f3), changes smoothly without exhibiting any singularities, which we refer to as \textit{crossovers}.
In the crossover region, we do not find any signal of the novel AFM structure with a longer period~\cite{Yu_2010, Yamada_2013, Timirgazin_2016}.
Due to the lack of a quantum field analysis, we cannot exclude the possibility of a first-order AMI-SI transition in the TDL.

\subsection{Continuous transitions and crossovers}\label{ContinuousTransitionsCrossovers}
Determining the transition points for continuous transitions is challenging.
For instance, the MF results shown in Fig.~\ref{fig:fig6}(g1-g3) suggest the continuous emergence of finite stripy magnetization $M^{(\text{s})}$ at the NM\textsubscript{3}-SI transition points, located at $U \approx 3.15$ and $3.51$ along the cutting lines $t^\prime=0.75$ and $0.8$ (not shown), respectively, for the $8$A torus.
Consistently, the entanglement entropy and stripy magnetization $M^{(\text{s})}$ curves, obtained from the DMRG calculations, do not demonstrate any discontinuities.
For $t^\prime < 0.8$, we identify the transition point by locating the maximum of the entanglement entropy, which results in an unclosed semicircular boundary for the NM\textsubscript{3} phase in the ground-state phase diagram, as shown in Fig.~\ref{fig:fig3}(d1) for $8$A.
For $t^\prime > 0.8$, the NM\textsubscript{3}-SI transition becomes a crossover, marked by the dashed line.

When NM\textsubscript{3} and SI are sandwiched with SM\textsubscript{1} for the $16$B and $24$B tori, the situation becomes quite complex, as illustrated in Figs.~\ref{fig:fig3}(d2,d3).
Here, we consider the transition from the SM\textsubscript{1} phase to the SI phase as an example.
In the MF scenario, jumps are observed, as shown in Fig.~\ref{fig:fig6}(h1).
For the $16$B torus, the jump corresponds to a first-order transition.
For the $24$B torus, the jump results from the exchange between two distinct stripy patterns, with alternating spin stripes along either the $x$ or $y$-axis.
However, in the DMRG results presented in Figs.~\ref{fig:fig6}(h2,h3), these jumps are replaced with shoulders, indicating smooth crossovers.

Moreover, in the ground-state phase diagrams shown in Figs.~\ref{fig:fig3}(c1-c3,d1-d3), we suggest that the AMM\textsubscript{1}-NM\textsubscript{3}, AMM\textsubscript{1}-SI, NM\textsubscript{3}-SI transitions may not exist in the TDL.

\subsection{VBS}\label{VBS}

\begin{figure}[t!]
\centering
\includegraphics[width=\linewidth]{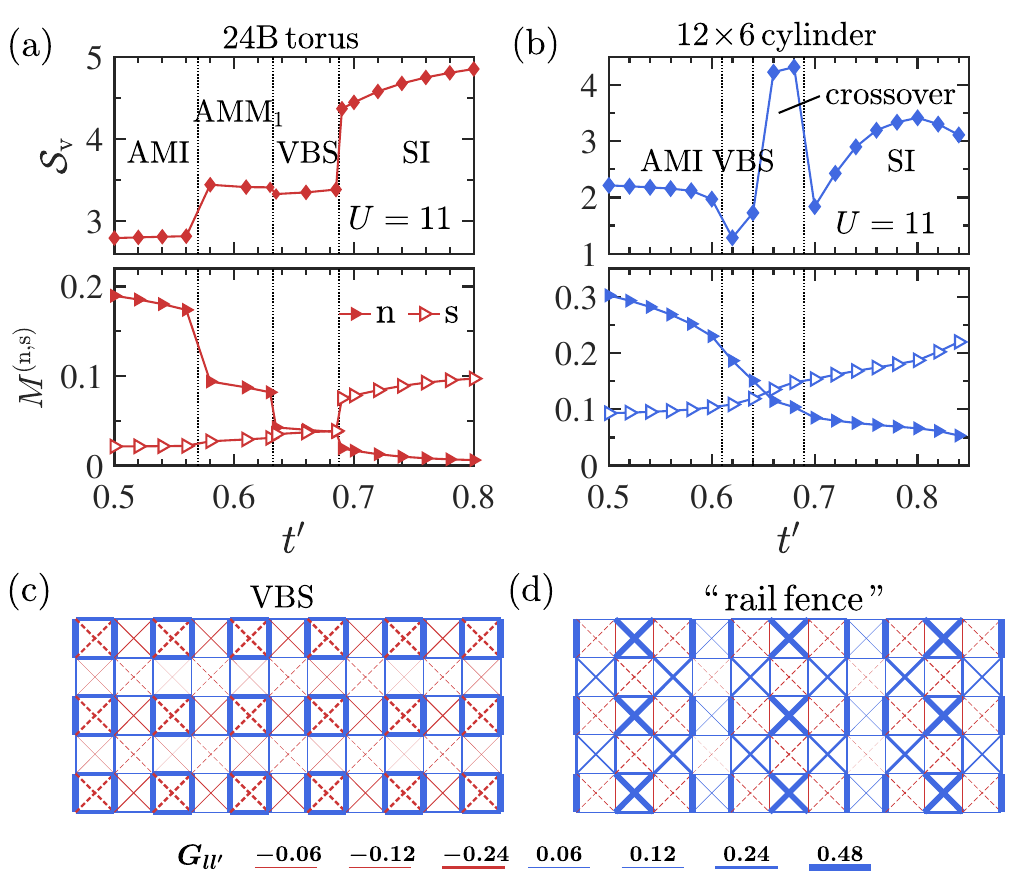}
\caption{The entanglement entropy $\mathcal{S}_\text{v}$ (\tcr{$\sblacklozenge[0.8]$}), and the magnetizations $M^{(\text{n})}$ (\tcr{$\sblacktriangleright[0.9]$}) and $M^{(\text{s})}$ (\tcr{$\svartriangleright[0.9]$}) as a function of the hopping amplitude $t^\prime$ for (a) a torus $24$B, and (b) a cylinder with $L_x \times L_y = 12 \times 6$.
On the cylinder, the distributions of NN and NNN spin correlations are shown for (c) $t^\prime=0.62$ and (d) $0.66$.
The red and blue lines indicate the ferromagnetic (FM) and AFM spin correlations, respectively.
We set the truncated bond dimensions $\chi=8,192$ and use $U=11$.
}
\label{fig:fig7}
\end{figure}

We identify an intermediate phase that emerges between the AMM\textsubscript{1} and SI phases, as shown in Fig.~\ref{fig:fig3}(d3).
For instance, along the cutting line $U=11$ for the $24$B torus, as indicated by jumps in the entanglement entropy and magnetizations in Fig.~\ref{fig:fig7}(a), the entire parameter region is divided into four distinct parts: an AMI phase, an AMM\textsubscript{1} phase, followed by an intermediate phase and an SI phase.
Notably, in the intermediate region $t^\prime \in [0.6325,\, 0.6875]$, both the staggered and stripy magnetizations are relatively small, implying that the intermediate phase may not be characterized by either the N\'{e}el or stripe orders.

To pin down the nature of the intermediate phase, we investigate a $12 \times 6$ cylinder, as shown in Figs.~\ref{fig:fig7}(b-d), with open boundary conditions along the $x$-axis and periodic boundary conditions along the $y$-axis.
First, we observe that the AMM\textsubscript{1} phase disappears at $U=11$, in comparison to the $24$B torus, as illustrated in Figs.~\ref{fig:fig7}(a,b).
This suggests that the phase boundary of AMM\textsubscript{1} strongly depends on the size and geometry of the selected cluster.
Next, for $t^\prime=0.62$ within the intermediate region $t^\prime \in [0.61,\, 0.64]$, which is close to the AMI phase region, we observe characteristic features on the cylinder: strong spin valence bonds are arranged in the $t_-$-plaquettes, while spin correlations on the $t_+$-plaquettes are extremely small, as shown in Fig.~\ref{fig:fig7}(c).
According to the spin correlation distribution, we attribute this intermediate region to the VBS phase, a potential candidate for the ground state of the model~\eqref{eq:model} in large $U$ region~\cite{Sindzingre_2002, Brenig_2004, Moukouri_2008, Chan_2011}.
After $t^\prime=0.69$, the ground state transitions to the stripe phase.
Since the stripe phase is non-collinear and shares the same point group symmetry as the VBS phase, the ground state undergoes a crossover from the VBS phase to the stripe phase.
As $t^\prime$ increases, the characteristic stripe patterns on the $t_+$-plaquettes, which feature strong NNN AFM links and weak NN FM links, gradually replace the VBS patterns.
Consequently, in the crossover region, we observe mixing patterns, such as the ``rail fence" shown in Fig.~\ref{fig:fig7}(d).

\section{Summary and conclusions}\label{sec:Conclusions}
We have investigated the ground-state phase diagram of $t$-$t^\prime$-$\delta$ model at half filling using both the MF and DMRG methods.
At an intermediate $\delta =0.5$, our MF analysis reveals a rich ground-state phase diagram in the thermodynamic limit, which includes three NM phases, two AMM phases, one AMI phase, two SM phases, and one SI phase.
We further discussed their magnetic properties, with a particular focus on altermagnetism, which is characterized by a mismatch in the topology of distinct Fermi surfaces for distinct species.
For each small selected torus, we have systematically demonstrated the agreements between the MF and DMRG results in terms of phase properties, phase transitions and crossovers.
Remarkably, for the large $24$B torus, we have identified a VBS phase in an intermediate-$t^\prime$ region where $U$ becomes dominant, a phase never observed in the $t$-$t^\prime$-$\delta$ model before.
Additionally, we have found VBS patterns with well-organized strong NN AFM bonds on the $t_-$-plaquettes in cylindrical geometries.
Compared to previous studies, our work presents numerical results for this model in an intermediate-$\delta$ region, complementing and extending beyond MF analysis, which demonstrates the close connection between altermagnetism and the intricate interplay between charge and spin orders.

\begin{acknowledgments}
We thank Lei Wang for the helpful discussions.
We acknowledge funding support from the Ministry of Science and Technology of the People's Republic of China (Grant No.~2022YFA1402700), the National Natural Science Foundation of China (Grants No.~U2230402, 12274187, 12174020, and 12247101) and the Fundamental Research Funds for the Central Universities (Grant No. lzujbky-2024-jdzx06), the Natural Science Foundation of Gansu Province (No. 22JR5RA389).
\end{acknowledgments}

\appendix
\label{APPENDIX}

\renewcommand\thefigure{S\arabic{figure}}

\setcounter{figure}{0}

\section{Mean-field ground-state phase diagram for $\delta=0.2$ and $0.8$}\label{app:appA}
\begin{figure}[h!]
\centering
\includegraphics[width=\columnwidth]{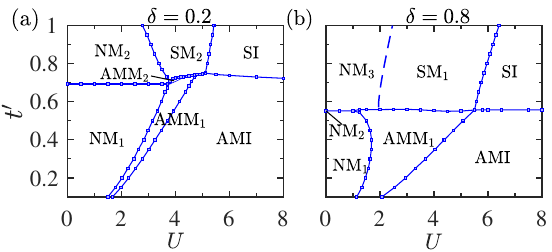}\\
\caption{The MF ground-state phase diagram in the TDL in the ($U$, $t^\prime$)-plane for (a) $\delta=0.2$, and (b) $0.8$.
}
\label{fig:figs1}
\end{figure}

In addition to $\delta=0.5$ mainly discussed in the main text, we also plot the MF groundstate phase diagram when $\delta=0.2$ and $0.8$ in Fig.~\ref{fig:figs1}.
Associated with Fig.~\ref{fig:fig2}(d) for $\delta=0.5$, we find some features:
(\textbf{1}) For small $t'$, the metallic phase NM\textsubscript{1} remains stable. For large $t'$, NM\textsubscript{3} remains stable. In contrast, NM\textsubscript{2} only survives at the transition point $t^\prime=t^\prime_\text{c}$ once $\delta$ exceed a threshold $\delta_\text{c}$.
(\textbf{2}) In the vicinity of $\delta=0.5$, four magnetic metallic phases, namely AMM\textsubscript{1}, AMM\textsubscript{2}, SM\textsubscript{1}, and SM\textsubscript{2}, emerge due to the strong competition between distinct energy scales in the model~\eqref{eq:model}. However, apart from $\delta=0.5$, some of these phases are absent.

\section{Entanglement entropy for other geometries}\label{app:appB}
\begin{figure}[h!]
\centering
\vskip0.1in
\includegraphics[width=0.7\columnwidth]{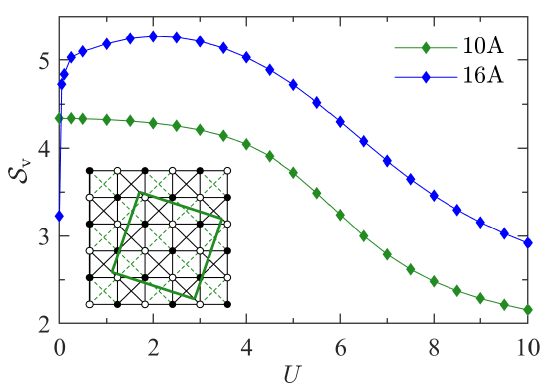}
\caption{The entanglement entropy $\mathcal{S}_\text{v}$ as a function of the on-site repulsion strength $U$ at $t^\prime =0.3$ and $\delta=0.5$ for the $10$A torus (inset), and the $16$A torus defined in Ref.~\cite{Maier_2005}.
We use the truncated bond dimensions of $\chi = 10,000$ in the U($1$)$\times$U($1$) DMRG calculations.}
\label{fig:figs2}
\end{figure}

We make benchmarks for other $10$A and $16$A tori~\cite{Maier_2005}, which cannot hit all high-symmetry momentum points in the FBZ.
Differently from $8$A, $16$B and $24$B discussed in the main text, we find that no jumps are observed in the curves of the entanglement entropy in Fig.~\ref{fig:figs2}.
Notably, at $U=0$ using the U($1$)$\times$U($1$) DMRG, a sudden reduction of the entanglement entropy happens due to the breaking of the chaotic $6$-fold degeneracy in the Hilbert space of $S^z = 0$, which is irrelevant to the presence of phase transitions.

\bibliography{ref}

\end{document}